\documentclass[prl,reprint,amsmath,amssymb,twocolumn,showpacs,letterpaper,superscriptaddress,10pt]{revtex4-1}
\usepackage{graphicx}
\begin{document}
\title{Magnetic field tuned superconductor-to-insulator transition at the LaAlO$_3$/SrTiO$_3$ interface}

\author{M. M. Mehta}
\affiliation{Department of Physics and Astronomy, Northwestern University, Evanston, IL 60208, USA}
\author{D.A. Dikin}
\affiliation{Department of Physics and Astronomy, Northwestern University, Evanston, IL 60208, USA}
\author{C.W. Bark} 
\affiliation{Department of Materials Science and Engineering, University of Wisconsin-Madison, Madison, WI 53706, USA}
\author{S. Ryu} 
\affiliation{Department of Materials Science and Engineering, University of Wisconsin-Madison, Madison, WI 53706, USA}
\author{C.M. Folkman} 
\affiliation{Department of Materials Science and Engineering, University of Wisconsin-Madison, Madison, WI 53706, USA}
\author{C.B. Eom} 
\affiliation{Department of Materials Science and Engineering, University of Wisconsin-Madison, Madison, WI 53706, USA}
\author{V. Chandrasekhar}
\email{v-chandrasekhar@northwestern.edu}
\affiliation{Department of Physics and Astronomy, Northwestern University, Evanston, IL 60208, USA}

\begin{abstract}

We present a study of the magnetic field tuned superconductor-to-insulator transition (SIT) in the electron gas that forms at the LaAlO$_3$/SrTiO$_3$ interface. We find that the magnetic field induces a transition into a weakly insulating state, as is observed for the electrostatically tuned SIT at this interface. Finite size scaling of the magnetoresistance yields the critical exponent product $z\nu \simeq$ 7/3, indicating that the transition is governed by quantum percolation effects. While such critical exponents have been reported previously for high resistance films, they have not been reported for a low resistance system like ours, with a maximum sheet resistance of $\approx$ 1.5 k$\Omega$, much less than the quantum of resistance  $R_Q \equiv h/4e^2 = 6.45$ k$\Omega$. 

\end{abstract}


\maketitle 

Two dimensional superconductors are known to undergo a superconductor-to-insulator transition (SIT) upon varying an external parameter like disorder or magnetic field \cite{fisher2,fisher}. This SIT is one of the most widely studied experimental realizations of a quantum phase transition (QPT) \cite{goldman,hebard,yazdani,ketterson,baturina}. Theoretically, the system is believed to transition into an insulating state beyond a certain critical value of the external tuning parameter. Two possible scenarios for the nature of this insulating state have been proposed: (i) the system is a Bose insulator, characterized by localized Cooper pairs with a non-zero pair amplitude \cite{fisher2,fisher}; or (ii) the system is a Fermi insulator, characterized by localized electrons and a vanishing pair amplitude signifying the complete destruction of superconductivity \cite{finkelstein}. Experimentally this transition has been studied on a variety of thin superconducting films prepared from different materials \cite{
goldman,hebard, yazdani,ketterson,baturina}. One of the observed characteristic signatures of the phase transition is the fan shaped curves of the temperature dependence of resistance, as a function of the disorder or the magnetic field, marking the transition from a zero resistance superconducting state to a high resistance insulating state. However, the strength and nature of the insulating state reached is different for different materials, and there is still debate about the universality of this transition as predicted by the theory. 

At finite temperatures $T$, one of the ways in which the SIT is studied is through the scaling of the resistance of the system with magnetic field $H$ near a critical point. According to theory, the resistance of the system takes the scaling form \cite{fisher},

\begin{equation}
 R(H,T) = R_Cf[(H - H_C)/T^{1/z\nu}]
 \label{eqn1}
\end{equation}
where $f$ is a function of $H$ and $T$.  Here $H_C$ is the critical field at which the transition occurs, $z$ is the dynamical critical exponent, $\nu$ is the correlation length exponent, and $R_C$ is a constant. Near the transition, the physics is governed by the competition between two length scales, the quantum fluctuations correlation length, $\xi \sim |H-H_C|^{-\nu}$, and a thermal length, $L_T \propto T^{1/z}$. The thermal length $L_T$ is the length scale at which the quantum fluctuations at finite frequency $\hbar \Omega$ are cut off by the temperature $k_BT$. The resistance is some universal function of the ratio of these two length scales which gives rise to the scaling form above. Such scaling behaviour is a strong indicator of the occurrence of a phase transition, especially in the case of a continuous phase transition such as the magnetic field tuned SIT, with the values of the critical exponents determining the universality class of the transition.

Here we report measurements of the perpendicular magnetic field tuned SIT that occurs in the two dimensional electron gas (2DEG) which forms at the interface between the two band insulators LaAlO$_3$ (LAO) and SrTiO$_3$ (STO) \cite{ohtomo}. It has been shown previously that this system can be tuned through a SIT on the application of a gate voltage \cite{caviglia,dikin}, which has the effect of changing the carrier density at the interface. Electrical transport measurements reported by various groups so far suggest that the system transitions into a weakly insulating state in the gate tuned SIT \cite{caviglia,dikin,mehta,schneider}. In this paper, we study the effect of a magnetic field on this system at various gate voltages at which the system is superconducting, with the normal-state sheet resistance of the 2DEG varying from $R^{N}_{\square} \approx$ 800 $\Omega$ to $R^{N}_{\square} \approx$ 1500 $\Omega$ for the range of measured gate voltages. We find a similar weakly insulating state in the magnetic 
field tuned transition at this interface. The resistance scales with $H$ and $T$ in accordance with Eq.  (\ref{eqn1}) with the critical exponent product $z\nu = 7/3$ giving the best scaling of the magnetoresistance (MR) data for all gate voltages. This indicates that quantum percolation effects dominate transport at the transition, i.e., Cooper pair tunneling (as opposed to direct transport) across a non-superconducting barrier gives rise to conduction as one approaches the transition from the insulating side \cite{dubi}. The system can be modeled as a network of superconducting grains (gap, $\Delta >$ 0) embedded in a non-superconducting material ($\Delta =$ 0) with strong $e-e$ interactions. This model is supported by the growing evidence for local inhomogeneity at the interface \cite{mehta,bert,pavlenko}.

In addition to superconductivity, this system also shows signature of ferromagnetism that coexists with the superconductivity \cite{dikin,bert,li}, where the ferromagnetic moment lies in the plane of the 2DEG \cite{bert,michaeli}. Due to the effect of the magnetization dynamics of the ferromagnet, the magnetoresistance of the system is hysteretic \cite{dikin,mehta}. In order to study the magnetic field-tuned transition at the interface, we have chosen to eliminate the effect of hysteresis due to the ferromagnet on the 2DEG by applying a small parallel field, $H_{\parallel} =$ 100 mT that aligns the moment of the ferromagnet in one direction. The sample growth and electrical transport properties have been discussed in earlier publications \cite{park,bark,dikin,mehta}. The electrical transport measurements of the sample were made using a standard a.c. lockin technique, with an excitation current of 19.2 nA and a frequency of 22.6 Hz in a dilution refrigerator.

\begin{figure}[ht]
\includegraphics[width=7cm]{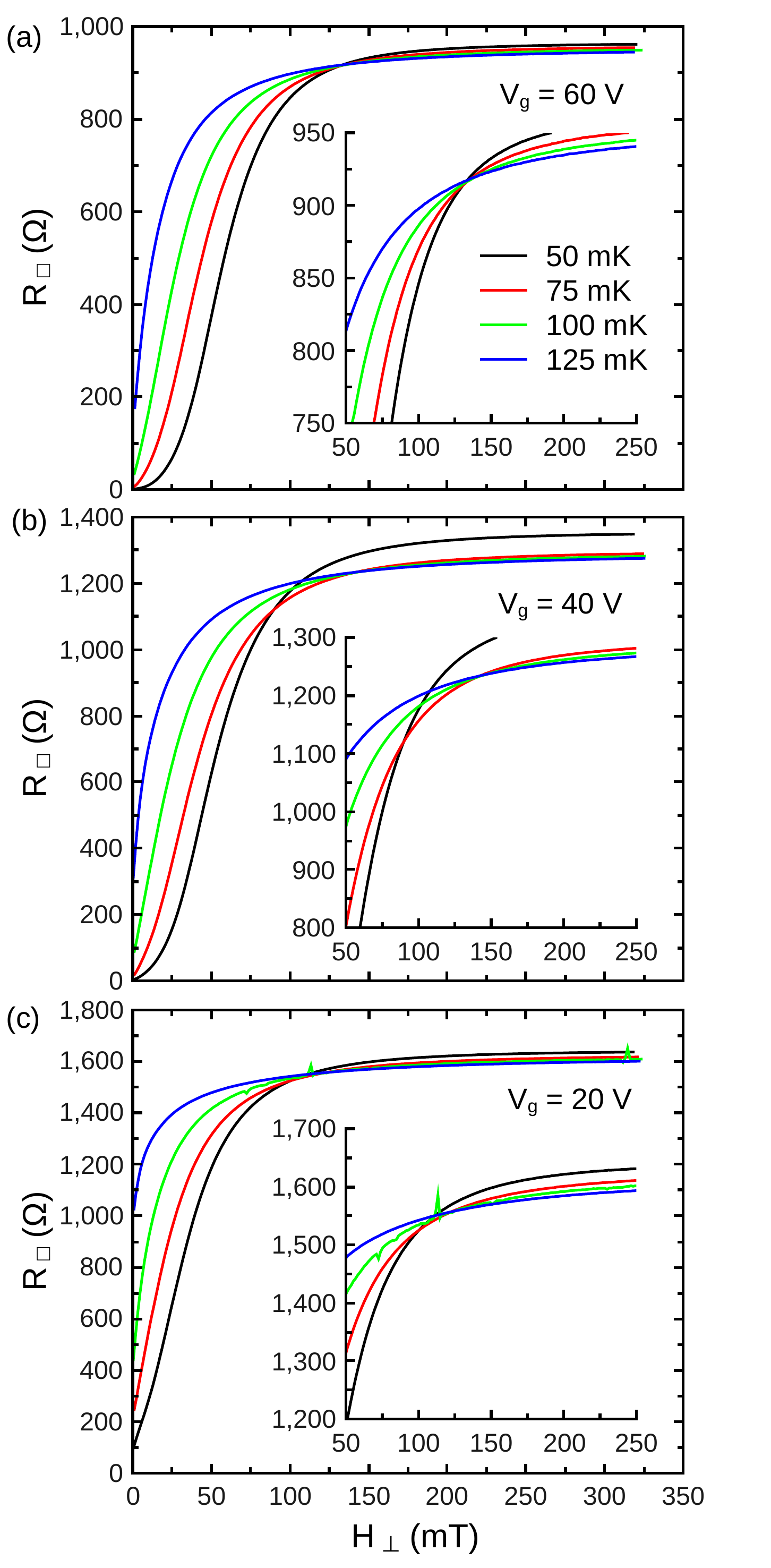}
\caption{(a), (b), and (c) are the MR at four different temperatures for the three different gate voltages, $V_g =$ 60 V, 40 V, and 20 V respectively, in order of decreasing $T_C$s. All the MR data are at temperatures which are lower than the measured $T_C$s at all the gate voltages. The insets show the zoomed in region of the MR where the different isotherms cross each other. Nominally, the field at which the crossing occurs is close to the critical field $H_C$ of the field induced SIT. }
\label{fig1}
\end{figure}

Each panel in Fig. 1 shows the MR of the system at four temperatures below the measured transition temperature $T_C$ of the sample, where we define $T_C$ as the midpoint of the resistive transition. The three panels show the MR for three different gate voltages at which the system is superconducting. Due to the application of a constant parallel field, the MR is non-hysteretic.  Hence, we show data for one sweep direction of $H_{\perp}$ ($H_{\perp}$ swept from 0 $\rightarrow$ 320 mT). (The parallel field also results in a small ($\sim$ 3 mT) asymmetry about $H_{\perp} =$ 0 mT due to the misalignment of the sample in the field.)  In these samples, the maximum $T_C$ occurs at $V_g =$ 60 V, with the $T_C$ decreasing upon further reduction of $V_g$. For all gate voltages, the system reaches a state of higher resistance with a weak temperature dependence upon the application of magnetic field.  It has been argued by us \cite{mehta} and others \cite{schneider} 
that the nature of the insulating state in the gate voltage tuned SIT can be 
described by taking into account the 
$e-e$ interactions in the electron gas. This picture describes the observed weak temperature dependence of the resistance in the electrostatically tuned SIT. An obvious question arises about the nature and strength of the insulating state in the magnetic field tuned transition, which we address below. 

It is seen from Fig. 1 that beyond a certain critical field (which is roughly the field at which the different MR isotherms cross; more on the determination of the critical field below) the MR increases slowly with the field at all temperatures for all the gate voltages. Unlike more resistive systems in which the field-tuned transition is studied \cite{steiner2,sambandamurthy,baturina}, we do not observe a large peak in the MR, at least in the measured field range. This peak in MR is believed to arise due to the localization of Cooper pairs in the insulating state, with the size of the peak increasing with decreasing temperature. One way to think about the origin of the peak is that an increasing magnetic field causes phase randomization between the neighbouring superconducting grains, which makes it difficult for Cooper pairs to tunnel between them, resulting in an increase in resistance. An observation of the MR peak is one of the signatures of the system being in the so-called Bose insulator phase. The 
fact that we do not observe this peak, coupled with the relatively small temperature dependence of the MR might suggest that the insulating state reached in the field tuned transition is dominated by Fermion physics, i.e., that the superconducting gap is uniformly destroyed throughout the sample and that the transition is more like a superconductor-to-metal transition. However, as mentioned above, the weakly insulating regime in the gate tuned SIT can be explained by the $e-e$ interactions in the system, and the distinction, if any, between a \textit{metal} with strong $e-e$ interactions and a weak insulator is difficult to resolve through the MR data. One therefore needs to perform a scaling analysis of the SIT, and from the critical exponents extracted, more can be said about the nature of the transition, as we show below.

In order to verify that the magnetic field does indeed cause a transition to a weakly insulating state, we have measured the differential resistance of the sample as a function of applied dc current $I_{dc}$ at various discrete magnetic fields. This is shown in Fig. 2 for $V_g =$ 60 V and $T $= 50 mK. As the field is increased, the system transitions from a superconducting to an insulating regime, as seen by the disappearance of the two dips in resistance at low bias between $H =$ 150 mT and $H =$ 159 mT.  For the highest measured field, $H_{\perp} = $ 318 mT, the fractional change in resistance between $I_{dc} =$ 0 $\mu$A and $I_{dc} =$ 1 $\mu$A is $\sim$ 3.2 $\%$. A similar change of resistance is observed in the insulating state reached by the gate voltage tuned transition \cite{mehta,schneider}. 

\begin{figure}[t]
\includegraphics[width=7cm]{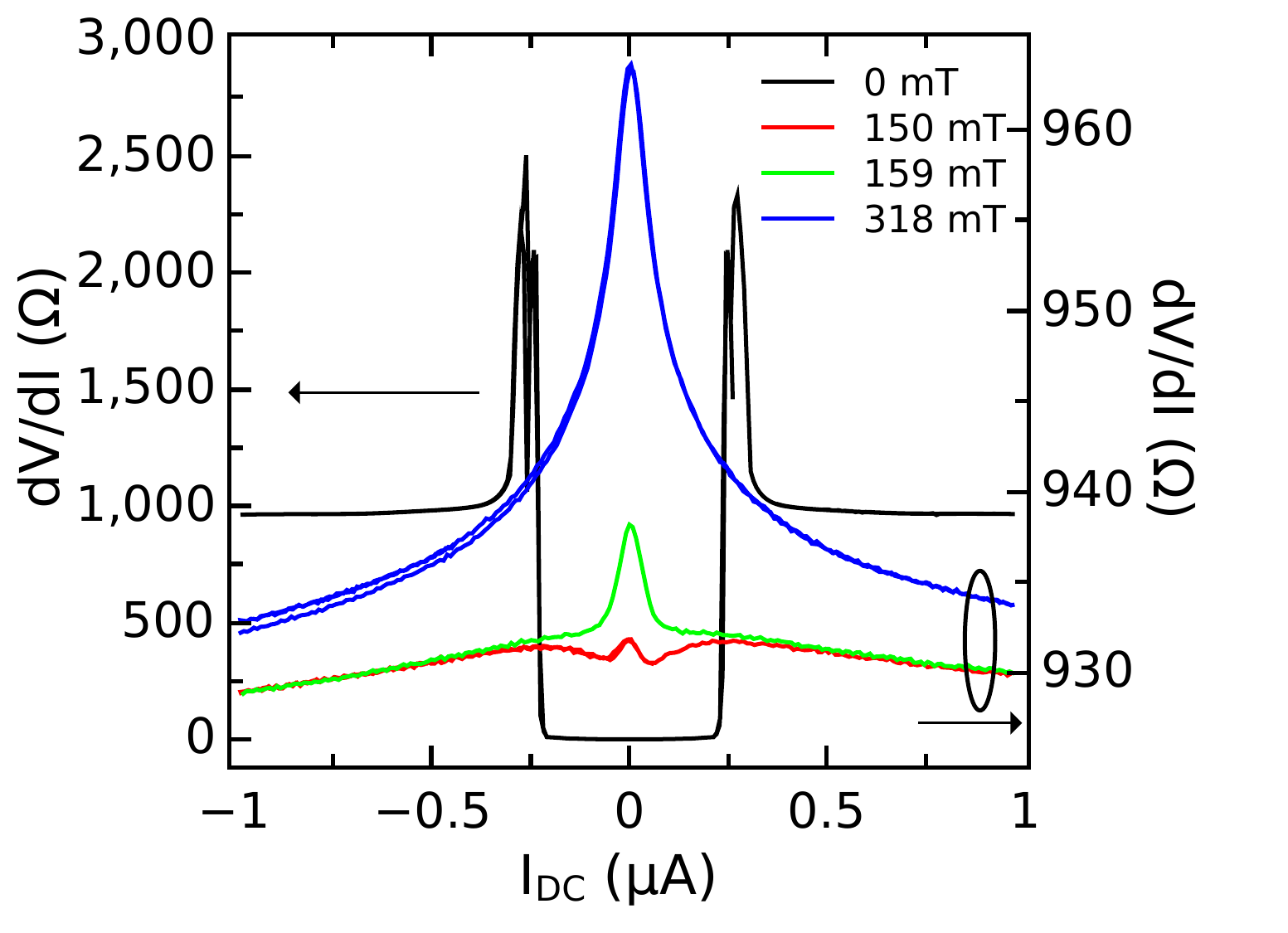}
\caption{The magnetic field induces a transition into a weakly insulating state as seen for $V_g$ = 60 V here. The transition happens close to $H_{\perp} =$ 159 mT. $T =$ 50 mK.}
\label{fig2}
\end{figure}

As discussed above, one of the ways in which the field tuned transition is analyzed is through the universal scaling behaviour of some physical property of the sample, such as resistance. A critical factor determining the successful scaling of experimental data is the determination of the critical field $H_C$. One way to determine $H_C$ is to see at what field the different MR isotherms cross (Fig. 1). However, this does not give an accurate measure of the $H_C$, especially for low resistance samples like ours. From Fig. 2 it can be seen that the transition happens close to $H_{\perp} =$ 159 mT for $V_g =$ 60 V. However, the determination of 
$H_C$ by changing the field in discrete steps is a very time consuming process, particularly for a system with such a large phase space. 

\begin{figure}[!b]
\includegraphics[width=7cm]{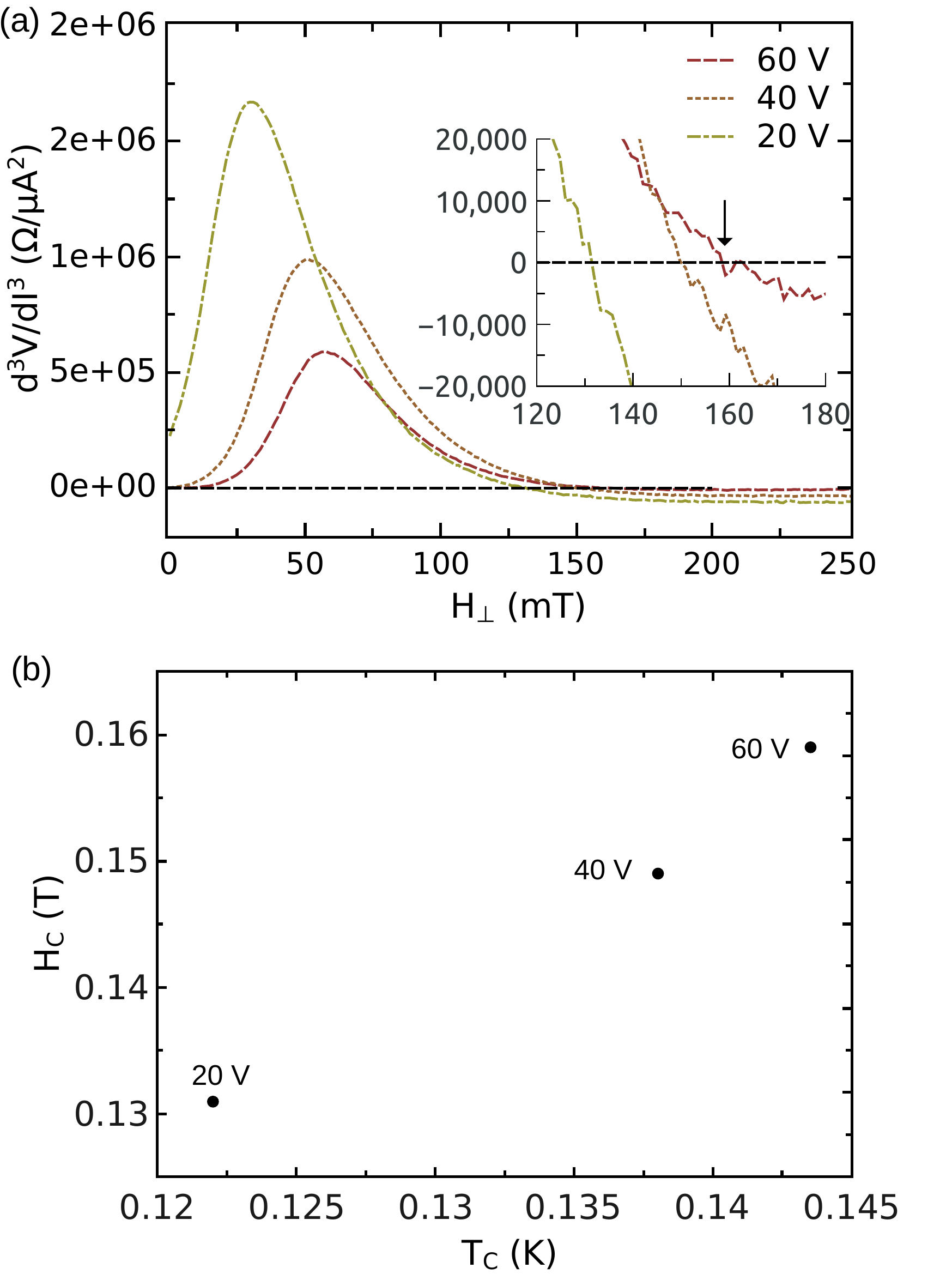}
\caption{(a) $d^3V/dI^3$ vs $H_{\perp}$ for different gate voltages. The horizontal dashed line indicates the zero 
of $d^3V/dI^3$. When the third derivative crosses this line as a function of the field, the system passes into an insulating state. 
 The inset shows the zoomed in region of the zero crossings for the three different gate voltages. The critical value of the field for $V_g$ = 60 V is indicated by the arrow, $H_C =$ 159 mT. $T =$ 50 mK. (b) Measured $H_C$ as a function of different measured $T_C$s for the different gate voltages. }
\label{fig3}
\end{figure}    

To ascertain the value of $H_C$ for the different gate voltages in the superconducting regime, we have measured the third derivative of voltage with current, $d^3V/dI^3$, as a function of the field. The advantage of this technique is that it is much faster to observe the field tuned transition compared to measuring the $R$ vs. $T$ or $I$-$V$ characteristics at discrete magnetic fields. When the system is not dc biased (i.e. $I_{dc} =$ 0), $dV/dI =$ 0 in the superconducting state, and it is positive and a  maximum in the insulating state (Fig. 2). Thus, it can be seen that $d^2V/dI^2 = $ 0 in both the superconducting and the insulating states. However, the third derivative, $d^3V/dI^3 \geq$ 0 in the superconducting regime, whereas $d^3V/dI^3 <$ 0 in the insulating regime. Therefore, the value of $H_{\perp}$ at which $d^3V/dI^3$ changes sign from positive to negative is the critical field $H_C$. 

Figure 3 shows $d^3V/dI^3$ vs $H_{\perp}$ for three different gate voltages at which the system is superconducting at zero field. Some non-monotonic features are seen at low fields, which arise due to the curvature in the shape of the MR (Fig. 1). As expected, the system transitions into the insulating state for all the gate voltages at higher fields, as seen by the change in sign of the curves, and the critical fields are different for the different gate voltages. For $V_g$ = 60 V, $H_C$ = 159 mT, indicated by the arrow. This value of $H_C$ is confirmed by the measurements of Fig. 2, where precisely at $H =$ 159 mT one sees the change from superconducting to insulating behaviour. Figure 3b shows the dependence of $H_C$ on the measured transition temperature, $T_C$, for the different gate voltages. It was argued by Fisher \cite{fisher} and demonstrated by Hebard and Paalanen \cite{hebard}, that near the critical point of the disorder driven transition, $H_C \sim T_C^{2/z}$. We do not see such behaviour in 
our data, but that could be because at the measured gate voltages we are still relatively far from the gate voltage at which the SIT occurs in the gate-voltage tuned transition. 

\begin{figure}[!ht]
\includegraphics[width=7cm]{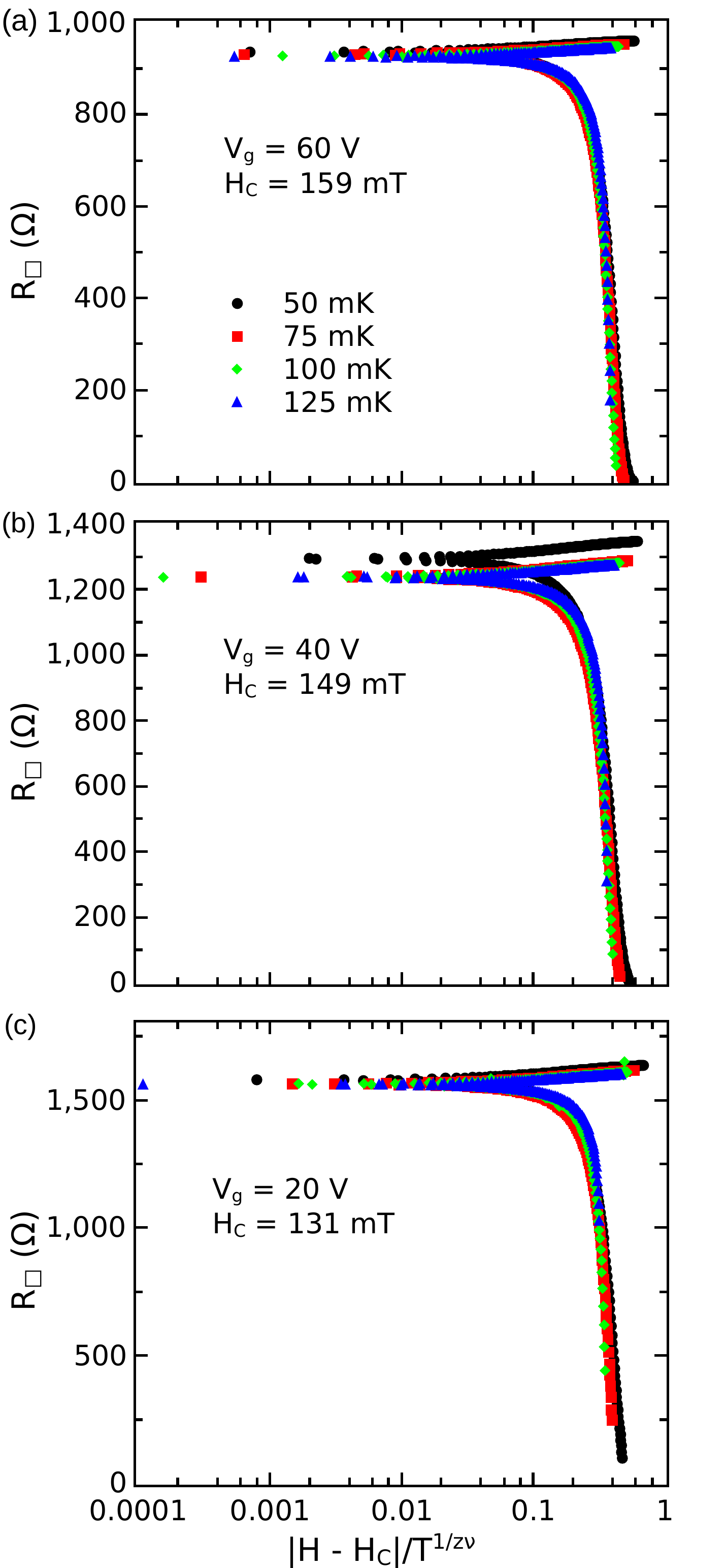}
\caption{$R_{\square}$ vs $|H-H_C|/T^{1/z\nu}$ for (a) $V_g =$ 60 V, (b) $V_g =$ 40 V, and (c) $V_g =$ 20 V. The critical exponent product $z\nu =$ 7/3 which
is predicted theoretically to occur for quantum percolation best fits the data, such that visually all the curves collapse onto each other. $H_C$ is obtained from third derivative measurements as explained in the text. }
\label{fig4}
\end{figure}  

With the values of $H_C$ extracted from the $d^3V/dI^3$ measurements one can proceed to the scaling analysis of the MR data. This is shown in Fig. 4 for the different gate voltages. The MR data are reasonably well scaled for the measured range of temperatures. The critical exponent product that best scales the data is  $z\nu =$  7/3, which is expected where quantum percolation is dominant \cite{dubi,steiner}. For inhomogeneous systems that show a phase transition from a conducting to a non-conducting state, quantum percolation signifies transport through charge tunneling (as opposed to direct transport) between two conducting regions across a non-conducting region. In the 2D SIT, the question is whether the principal charge transport is through Cooper pairs (Bosons) or through electrons (Fermions) as the magnetic field is increased.  Given that our samples have low disorder (minimum $k_F l \approx$ 19, at $V_g =$ 20 V; where $k_F$ is the Fermi wave vector and $l$ is the elastic mean free path), it is 
unlikely that electrons would be localized on isolated grains in the weakly insulating regime.  If the dominant carriers were 
Cooper pairs, however, one could imagine that the Cooper pairs would be localized on superconducting islands, isolated from Cooper pairs on other islands by intervening normal regions, with the principal transport mechanism being tunneling of Cooper pairs from one island to another.  Consequently, the dominant charge transport at the transition occurs through Cooper pair tunneling, implying that even beyond the transition into the insulating side, Cooper pairs and hence the superconducting order parameter survive locally while global phase coherence is lost. Of course, a more local probe (like a scanning probe microscope) of the superconducting phase correlations would help shed more light on the matter. 

In conclusion, we have measured the magnetic field induced superconductor-to-insulator transition at the LAO/STO interface. As seen in the gate voltage tuned transition at this interface, the magnetic field also causes the system to transition into a weakly insulating state. This state is characterized by a weak temperature dependence of the MR. The critical exponents obtained from the scaling of the MR data indicate that quantum percolation effects dominate the physics at this transition, signifying the tunneling of Cooper pairs contributes to the transport in this system on the insulating side. 

\begin{acknowledgments}
Work at Northwestern was supported by a grant from the DOE Office of Basic Energy Sciences under grant no. DE-FG02-06ER46346. Work at the University of Wisconsin was supported by funding from the DOE Office of Basic Energy Sciences under award number DE-FG02-06ER46327 and the National Science Foundation under grant no. DMR-1234096.
\end{acknowledgments}

\end{document}